\documentclass[%
reprint,
superscriptaddress,
 amsmath,amssymb,
 aip,
]{revtex4-2}

\usepackage{threeparttable}
\usepackage{graphicx}
\usepackage{dcolumn}
\usepackage{bm}
\usepackage{xcolor}

\begin{document}


\title{Femtosecond laser-induced sub-wavelength plasma inside dielectrics: I. Field enhancement}

\author{Kazem Ardaneh}
\email{kazem.arrdaneh@gmail.com}
\author{Remi Meyer}
\author{Mostafa Hassan}
\author{Remo Giust}
\author{Benoit Morel}
\affiliation{FEMTO-ST Institute, Univ. Bourgogne Franche-Comt\'e, CNRS,15B avenue des Montboucons,25030, Besan\c{c}on Cedex, France}
\author{Arnaud Couairon}
\affiliation{CPHT, CNRS, Ecole Polytechnique, Institut Polytechnique de Paris, Route de Saclay, F-91128 Palaiseau, France}
\author{Guy Bonnaud} 
\affiliation{CEA, Centre de Paris-Saclay, DRF, Univ.  Paris-Saclay, 91191 Gif-sur-Yvette, France}
\author{Francois Courvoisier}
\email{francois.courvoisier@femto-st.fr}
\affiliation{FEMTO-ST Institute, Univ. Bourgogne Franche-Comt\'e, CNRS,15B avenue des Montboucons,25030, Besan\c{c}on Cedex, France}%

\date{\today}

\begin{abstract}
The creation of high energy density ($\gtrsim10^6$ joules per cm$^3$) over-critical plasmas in a large volume has essential applications in the study of warm dense matter, being present in the hot cores of stars and planets.  It was recently shown that femtosecond Bessel beams enable creating over-critical plasmas inside sapphire with sub-wavelength radius and several tens of micrometers in length.  Here, the dependence of field structure and absorption mechanism on the plasma density transverse profile are investigated by performing self-consistent Particle-In-Cell (PIC) simulations.  Two { limiting} cases are considered: one is a homogeneous step-like profile, that can sustain plasmon formation, the second is an inhomogeneous Gaussian profile, where resonance absorption occurs.  Comparing experimental absorption measures to analytical predictions allows determining the plasma parameters used in PIC simulations. The PIC simulation results are in good agreement with experimental diagnostics of total absorption, near-field fluence distribution, and far-field radiation pattern. We show that in each case an ambipolar field forms at the plasma surface due to the expansion of the hot electrons and that electron sound waves propagate into the over-critical region.
\end{abstract}

\maketitle

\section {Introduction}\label{Introduction}
The focusing of a high-intensity ultrashort pulse on the surface of solids creates over-critical hot plasma via field and collisional ionization. The over-critical plasma is key to efficiently transfer the laser energy into the solid and is crucial for various applications, { e.g., plasma mirrors,\cite{Ziener_2003,Doumy_2004,thaury_2007} secondary radiation sources such as extreme ultraviolet (EUV) and X-ray sources.\cite{Teubner_1993,Sauerbrey_1994,Murnane_1989,Brambrink_2009} It is also crucial for fundamental investigations on generation of extreme pressures such as the one of warm dense matter that is the state of matter in the core of several astrophysical objects, e.g., planets and stars.\cite{Chabrier_2009, Helled_2010}} Particularly for the investigation of warm dense matter, it is desirable to create over-critical and hot plasmas over a large volume. 

Over-critical hot plasmas are easily created by tightly focusing a laser beam at the surface of solids. { However, the generated plasma expands into the vacuum. It would be attractive to create large volumes of plasmas within solids simultaneously at both very high temperatures and very high pressures; typically $>100$~eV and $>10^5$~J/cm$^3$.  The study of these plasmas is important for our understanding of high-energy-density physics where complex sub-fields of physics (quantum/classic, weakly/strongly coupled) intersect and predictions are very challenging.
In contrast to the surface of solids,  the generation of these plasmas inside solids using conventional Gaussian beams is restricted to small volumes due to plasma defocusing as discussed in Ref. \cite{ardaneh_2021}}

Non-diffracting femtosecond Bessel beams allow { creating plasmas with much larger volumes. These beams }are propagation-invariant solutions of Helmholtz equation, formed by a cylindrically-symmetric interference, with a transverse profile following the Bessel function of the first kind. \cite{Durnin87} They can sustain a subwavelength focal spot over a propagation distance not limited by the Rayleigh range. Subwavelength focus sustained over a centimeter scale has been demonstrated in Ref. \cite{Meyer2019}

 Single-shot femtosecond Bessel beams have been used to create sub-wavelength channels inside dielectrics such as fused silica and sapphire  \cite{Bhuyan_2010, Bhuyan2014,Rapp_2016, Bhuyan2017} and showed that tightly focusing a  single femtosecond pulse with ~\textmu J energy, shaped as a  Bessel Beam inside a dielectric, can create a 20~\textmu m channel with a diameter of about 400 nm. { In a Bessel beam, the reflection on a plasma rod at the center is superimposed to the transmitted beam. By measuring the transmitted energy, we could show that the conditions where a void channel can form were corresponding to a transmission drop by less than 50\%. This corresponds to} an absorption of more than 50\% of the laser energy indicating an energy density in the order of MJ/cm$^3$. The absorbed energy is a few orders of magnitude greater than the ionization energy required for creating a fully ionized plasma rod at the input laser critical density, $n_{\rm c}=1.7\times10^{21}\,{\rm cm^{-3}}$ for a 800 nm laser. Hence, the single-shot femtosecond Bessel beam allows creating over-critical plasma with sub-wavelength radius but over several tens of micrometers length and potentially over arbitrary distances. 

 \begin{figure*}[!htb]
\begin{center}
\includegraphics[width=\textwidth]{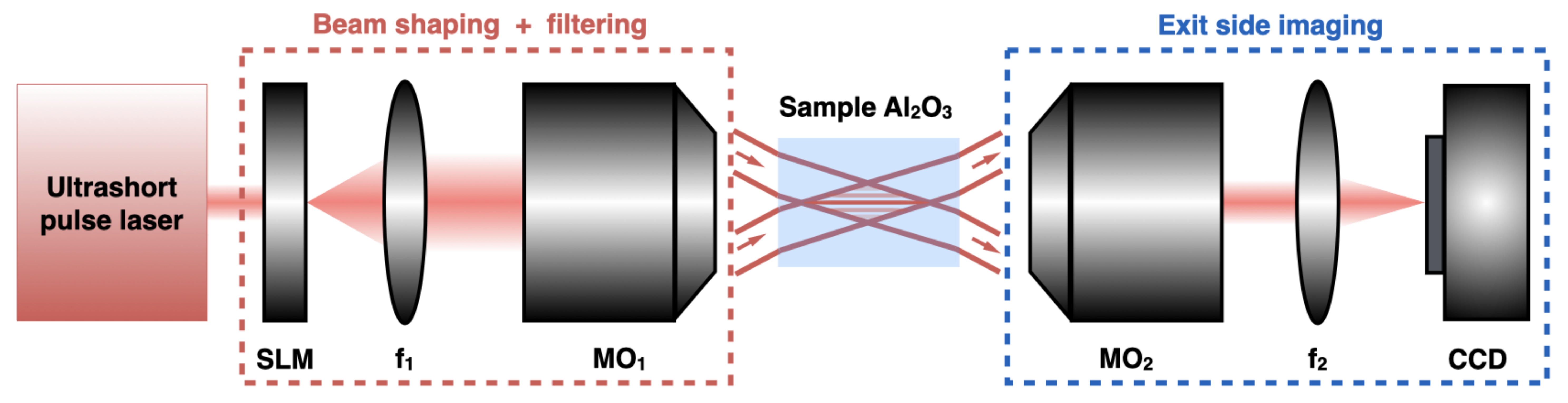}
\caption{{ Schematic of the experimental setup for femtosecond Bessel beam shaping and imaging of propagation in sapphire. Beam scanning allows reconstructing the fluence distribution inside the sample in three dimensions. Far-field scanning is achievable by changing lens f$_2$ to image the back focal plane of the second microscope objective onto the CCD. SLM: spatial light modulators, MO: microscope objective, CCD: charge-coupled device.}}
\label{exp_setup}
\end{center}
\end{figure*}

Here, we focus on the mechanism driving the high absorption that is experimentally observed and the dependence of the physics on the transverse density profile of the laser-induced plasma. Since direct measurement of plasma parameters inside dielectrics is not available using optical diagnostics, we focus on two { limiting} cases.
In a first case of a homogeneous and over-critical plasma, one can anticipate the formation of surface plasmon, an electromagnetic wave arising due to the permittivity difference between two neighboring materials.\cite{Raether_1988} In the second case, if the generated plasma is inhomogeneous (presence of a density gradient), the process of linear mode conversion can excite electron plasma waves from the $p-$polarized components of the Bessel beam.  The linear mode conversion can lead to an absorption of up to 70 percent.\cite{Denisov_1957,Pert_1978,kruer_1988} This process is electrostatic and relies on the resonance of plasma waves at the critical surface, where the plasma density reaches the critical density. Resonance absorption has been widely investigated for plasma at the surface of materials in the context of inertial confinement fusion. \cite{Milchberg1988,Murnane_1989,Kieffer1989,Fedosejevs_1990,Teubner_1993,Price_1995,Teubner_1996,Bastiani_1997}

To reproduce our experiments, we have performed first principle Particle-In-Cell (PIC) simulations using full 3D massively parallel EPOCH code.\cite{Arber_2015}. In contrast with the widely used two-temperature model, for instance used by Beuton {\it et al}\cite{Beuton_2021} for numerical study of dielectric modifications by femtosecond laser pulses, that simulate only collisional effects,  PIC simulations are best suited to capture collisionless effects, which account for most of the absorption as we have shown in the case of a Bessel-shaped pulse onto a plasma rod with an over-critical Gaussian density profile \cite{ardaneh_2021}. 

This paper is organized as follows. After briefly describing our experimental setup in Sec. \ref{Experiments}, we derive, in Sec. \ref{Theoretical constraints}, using simple theoretical models, the constraints on the plasma parameters. The numerical setups for PIC simulations are presented in Sec. \ref{Simulation setups}. In Sec. \ref{Results}, we compare our numerical results to experimental measurements and discuss the physics of the two { limiting} cases.  We shall see that while both { limiting} cases can yield with an adequate choice of plasma dimensions similarly good agreement with experimental results, the field structure and the resulting electron heating are basically different.

\section {Experiments}\label{Experiments}
Our experiments are based on generating a micro-Bessel beam inside c-cut sapphire, with a $14^{\circ}$ cone angle ($\theta=25^{\circ}$ in air), {\it i.e.}, the angle made by the optical rays with the optical axis. After the interaction, the Bessel beam is collected using a high numerical aperture microscope objective and imaged onto a camera. The procedures are detailed in Ref. \cite{Xie2015} and Ref. \cite{ardaneh_2021} { A simplified schematic of the experimental setup is shown in Fig. \ref{exp_setup}.}

Near-field imaging of the beam will be shown in Fig.~\ref{NearField}(a) together with numerical simulations results.  We have determined that the overall absorption factor is 50\% and that the absorption is localized in the region of the highest intensity of the central lobe ({\it i.e.}, not in the pre-focal region as it can occur with Gaussian beams).\cite{ardaneh_2021}

\section {Theoretical  constraints} \label{Theoretical constraints}
The absence of direct measurement techniques for plasma parameters inside the dielectrics strengthens the importance of the theoretical estimations of plasma parameters to perform numerical simulations. Here, we constrained the parameters of generated plasmas to lead to the same absorption as measurements using the conditions for surface plasmon formation and resonance absorption of short pulses. 

\subsection {Surface plasmon} \label{Surface plasmon}

Tightly focusing femtosecond laser pulses inside the dielectrics can lead to the nano-plasmas formation during the onset of the laser through field ionization. If the plasma density exceeds the critical density at the laser frequency, $n_{\rm c}(\omega)=\omega^{2}m_{\rm e}/ 4\pi e^{2}$, it provides proper conditions for surface plasmon excitation as discussed in several works.\cite{Joel_1981,Chen_1983,Boyd_1984,Rajeev_2003,Rajeev_2006,Rajeev_2007,Liao_15} To obtain a simple estimate of the plasma density and effective damping frequency necessary for an absorption factor of 50\%, we consider a planar geometry, {\it i.e.}, a two-layer system in the $xz-$plane, infinite in the $y-$direction, with permittivities $\epsilon_2$ for $x>0$, and $\epsilon_1$ for $x<0$, respectively. A $p-$polarized monochromatic laser interacts with this two-layer system. The wavevector is given by ${\mathbf{k}}_{0}={\mathbf{\hat{x}}} k_{\rm 0}\cos i + {\mathbf{\hat{z}}} k_{\rm 0}\sin i$ where $i$ is the angle measured from the normal to the interface. The magnetic field is along the $y-$direction, while the electric field is in the $xz-$plane. Applying the boundary conditions for the electric and magnetic fields in the interface of the two dielectrics, one can derive the wavevector along the $z-$, and $x-$directions as follows:\cite{Raether_1988}

\begin{subequations}
\label{SPE}
\begin{align}
k_{\rm z} &= k_0(\frac{\epsilon_1\epsilon_2}{\epsilon_1+\epsilon_2})^{1/2}\\
k_{\rm xi} &= \left(\epsilon_{\rm i}k_0^2-k^2_{\rm z}\right)^{1/2}
\end{align}
\end{subequations}

Considering $\epsilon_2 = 3.13$ (sapphire), $\Re\epsilon_1< 0$, and $|\Re\epsilon_1|>\epsilon_2$ (metal), one can see that $k_{\rm z} > k_0$, and $k_{\rm xi} < 0$. It means that the fields have their maxima at the interface $x = 0$, and decay in both sides of $x-$coordinate.   We used the Drude model for the metal permittivity in Eqs. (\ref{SPE}):

\begin{align}\label{eps_plasma}
\epsilon_{1}(\omega)=\epsilon_{2}-\frac{\omega_{\rm pe}^{2}/\omega^{2}}{1+j \nu_{\rm eff}/ \omega}
\end{align}

\noindent where $\omega_{\rm pe}=(4\pi n e^2/m_{\rm e})^{1/2}$ is the electron plasma frequency, and $\nu_{\rm eff}$ is the effective damping frequency in the metal. We constrain the electron density $n$ and $\nu_{\rm eff}$ using the absorption factor of a $p-$polarized laser:\cite{Reitz_2008}

\begin{equation}\label{indx1}
A_{\rm p}=1-\left|\frac{\sqrt{\epsilon_2} \cos i_1-\sqrt{\epsilon_1} \cos i_2}{\sqrt{\epsilon_2} \cos i_1+\sqrt{\epsilon_1} \cos i_2}\right|^{2}
\end{equation}

\begin{figure}[!htb]
\begin{center}
\includegraphics[width=\columnwidth]{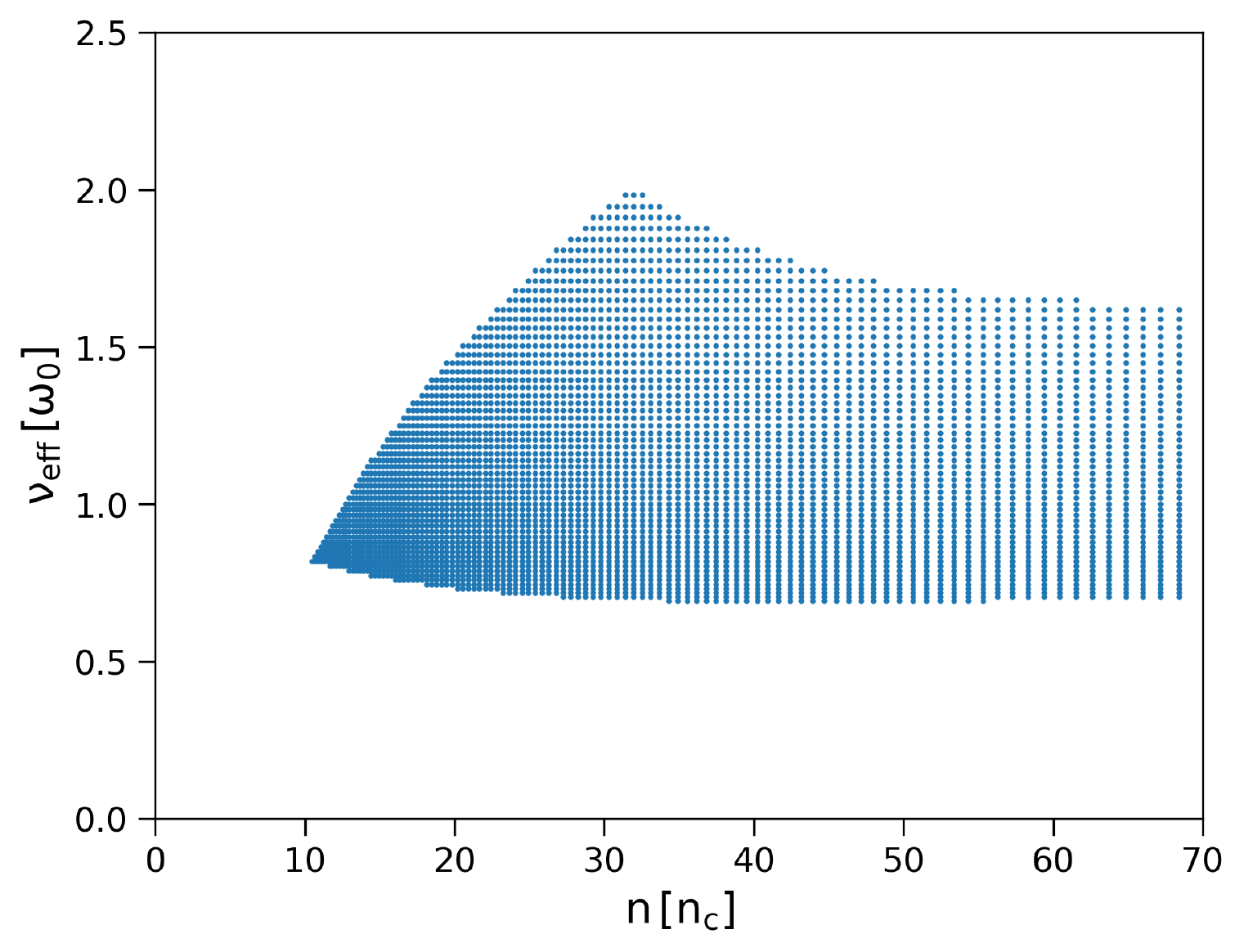}
\caption{Condition for surface plasmon excitation for an absorption of $0.5\leq A_{\rm p}\leq 0.7$. We determined the electron density and effective damping frequency by solving Eqs. (\ref{SPE}a) and (\ref{indx1}). Each point corresponds to one solution. }
\label{nu_tem}
\end{center}
\end{figure}

For an absorption of $0.5\leq A_{\rm p}\leq 0.7$, $\epsilon_{\rm 2}=3.13$, and $i_2=76^{\circ}$ ($i_2=\pi/2 -\theta$ for $\theta=14^{\circ}$ in sapphire), $i_1$ in plasma is  computed from Snell's law. We have determined $n$ and $\nu_{\rm eff}$ by solving Eqs. (\ref{SPE}a) and (\ref{indx1}) as shown in Fig. \ref{nu_tem}. The range of $\nu_{\rm eff}$ and $n$ for excitation of surface plasmons are respectively $0.7\omega_0\lesssim \nu_{\rm eff}\lesssim 2\omega_0$ (a damping time in the range of $0.2\,{\rm fs}\lesssim \tau_{\rm eff} \lesssim0.6\,{\rm fs}$), and $10n_{\rm c}\lesssim n\lesssim 70n_{\rm c}$.  The upper limit of the electron density is the atomic number density for the sapphire. The orders of magnitudes using this basic model of a monochromatic laser interacting with a plane interface are consistent with the values reported in the literature for ultrafast generation of plasma inside a transparent solid.\cite{Sun_2004}

\subsection {Resonance absorption} \label{Resonance absorption}

Hereabove the plasmon surface excitation was bound to a sharp density profile plasma. In contrast, in the case of an inhomogeneous plasma density, the presence of a gradient allows the excitation, by the laser pulse, of electron plasma waves (volume plasmons) along the density gradient, under $p-$polarized laser illumination of the plasmas. \cite{Denisov_1957,Pert_1978,kruer_1988} In this case, the dispersion relation defines the wavevector of the laser in the direction of the density gradient as $k_{\rm x}^2=k_0^2(\cos^2i-n/n_{\rm c})$ where $i$ is the incident angle measured relative to the density gradient. At the turning point defined by the density $n=n_{\rm c}\cos^2i$, the laser wave is reflected. The evanescent wave tunnels beyond the turning point and approaches the critical surface. The electric field component parallel to the density gradient creates charge separation due to the electron oscillations. The dielectric function vanishes at $n=n_{\rm c}$ and the driven electron plasma waves are in resonance. The electric field component parallel to the density gradient is given by:\cite{Denisov_1957,Pert_1978,kruer_1988}

\begin{equation}\label{RA_SP_1}
	E_{\rm x}=\frac{1}{\epsilon_1\sqrt{2\pi L\omega_0/c}}E_0(\omega_0)\Phi(q)
\end{equation}

where $q=(\omega_0 L/c)^{2/3}\sin^2i$ is the Denisov absorption parameter, $L$ the characteristic density gradient length, and function $\Phi$ describes the decay of the laser beyond the turning point and is expressed in terms of the Airy functions:\cite{Denisov_1957,Pert_1978}

\begin{equation}\label{RA_SP_2}
	\Phi(q)=\frac{4\pi^{1/2} q^{1/2}[{\rm Ai}(q)]^{3/2}}{|{\rm d}{\rm Ai}(q)/{\rm d}q|^{1/2}}
\end{equation}

One can extend the resonant absorption for a short pulse using the superposition principle. \cite{Nazarenko_1995,Palastro_2018} In the linear regime, each frequency of the incident pulse undergoes resonance absorption independently.  Hence, we can obtain the parallel electric field by integrating over all the frequencies, neglecting the small spatial variation of the critical frequency with the wavelength, $\delta L/L = 2\delta\omega/\omega\approx 0.02$ for pulses used in our experiments. 

\begin{equation}\label{RA_SP_6}
	E_{\rm x}=\int {\rm d}\omega F(\omega)E(\omega)
\end{equation}
where $F(\omega)=\Phi(\omega)/{\epsilon_1\sqrt{2\pi L\omega/c}}$ and $E(\omega)$ is spectral amplitude of the laser pulse. We assumed that the plasma density linearly increases with the $x$-coordinate ($n/n_{\rm c}=1+x/L$, the origin of coordinates is chosen so that with $x>0$ $\epsilon_{1}(x)<0$, and with $x<0$ $\epsilon_{1}(x)>0$). The dielectric function $\epsilon_1$ for the plasma then reads: 

\begin{equation}\label{RA_SP_4}
\epsilon_{1}(\omega,x)=\epsilon_{2}-\frac{\omega^{2}_0(1+x/L)/\omega^2}{1+j\nu_{\rm eff}/\omega}
\end{equation}

The absorption factor $A_{\rm p}$ is the ratio of the absorbed energy flux to the incident laser energy flux ($A_{\rm p}=I_{\rm A}/I_0$). The absorbed energy flux reads

\begin{equation}\label{RA_SP_7}
	I_{\rm A}=\int {\rm d}x\nu_{\rm eff} \frac{E_{\rm x}^2}{8\pi}
\end{equation}
The absorption factor is $A_{\rm p}=\Phi^2/2$ for the monochromatic laser wave in vacuum.\cite{Denisov_1957,kruer_1988} 

\begin{figure}[!htb]
\begin{center}
\includegraphics[width=\columnwidth]{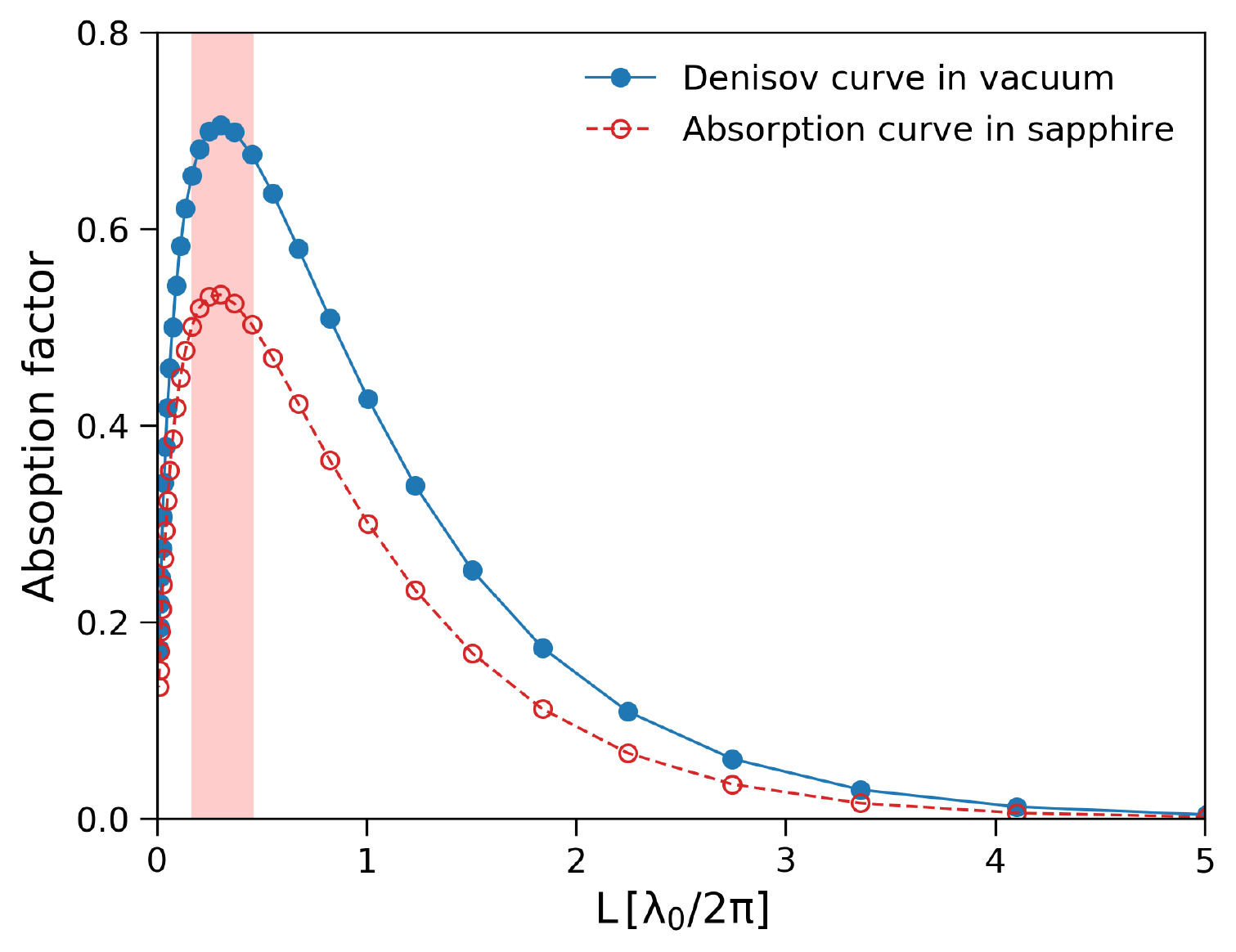}
\caption{Absorption factor as a function of plasma scale length for a ${\rm 100\,fs}$ Gaussian pulse (red dashed) and the Denisov absorption curve (blue solid). The red box shows the constraint on the plasma scale length for an absorption of $A_{\rm p}\gtrsim50$\%.}
\label{appendix3}
\end{center}
\end{figure}

In Fig. \ref{appendix3}, we have plotted the absorption factor as a function of plasma scale length $L$ for a Gaussian pulse with a central wavelength of $\lambda_0=800\, {\rm nm}$, a duration of ${\rm FWHM=100\,fs}$, an incident angle of $i=76^{\circ}$ (cone angle of $\theta=14^{\circ}$ in sapphire), $\epsilon_{\rm 2}=3.13$, and a effective damping frequency of $\nu_{\rm eff}=1.2\omega_0$ (Sec. \ref{Surface plasmon}).  The Denisov absorption curve is also shown for comparison. The difference between the two curves is due to the permittivity and effective damping frequency. One expects an absorption of about $\gtrsim50\%$ in a plasma with a scale length in the range of $L\approx 20-60\,{\rm nm}$.  For a maximum density of $10n_{\rm c}$ (Sec. \ref{Surface plasmon}), a scale length in the range of $L\approx 20-60\,{\rm nm}$  gives the plasma critical radius $R_{\rm c}$, the radius at which the density reaches the critical density, in the range of $\approx 180-540\,{\rm nm}$. 

As a summary for section~\ref{Theoretical constraints}, the constraints on plasma parameters for an absorption factor between $0.5-0.7$ as in our measurements are an electron density between $10-70n_{\rm c}$, a damping frequency between $0.7-2\omega_0$, and a critical radius within $180-540\,{\rm nm}$.

We have experimentally characterized, by { post-processing} imaging of the samples, that the plasma created in the Bessel beam configuration has an elliptical shape, oriented perpendicular to the polarization, {see Fig. 3(e) in Ref. \cite{ardaneh_2021}}
We can remark that the elliptical shape  can be understood if one considers a  circular dielectric rod of permittivity $\epsilon_{1}$ placed in an external homogeneous electric field in a medium of permittivity $\epsilon_{2}$.  For a sub-critical plasma, the electric field is amplified at the equatorial plane, perpendicular to the external field, in comparison with the polar plane, parallel to it, with a ratio of $\epsilon_{2}/\epsilon_{1}$.\cite{Reitz_2008} The field ionization is, therefore, more efficient in the equatorial plane and ionizes more the dielectric medium. It leads to an elliptical plasma rod elongated perpendicular to the laser polarization.  

\section {Simulation setups} \label {Simulation setups} 
We have performed self-consistent PIC simulations using the 3D massively parallel electromagnetic code EPOCH. \cite{Arber_2015} We have listed the parameters for the simulations in Table \ref{dsh}. The required energy for creating a plasma  with the volume of the experimental void formed in sapphire (typ. 0.4 ~\textmu m diameter over a length of 18~\textmu m), $\approx2.5$~\textmu m$^3$, at a critical density is in the order of nJ for ionization energy of $\approx 10\,{\rm eV}$. Therefore, we used a fully ionized plasma composed of electrons and ions. {To determine the number of free electrons per ion in ionization of sapphire, one needs to know the electron density of states (DOS). Another point is to know up to what energy gap the multiphoton ionization can promote electrons from the valence band to into the conduction band; the multiphoton cross-section depends on the laser frequency, polarization, number of absorbed photons, and the energy gap. We estimated DOS using data provided in Ref.\cite{ching_1994} We found that field ionization can quickly promote 1-2 electrons per molecule of sapphire from $\approx$ 0.5 eV below the valence band maximum to $\approx$ 3 eV above the conduction band minimum. In our simulation, we set the number of ions to be equal to the number of electrons, $Z=1$. Although the calculations become more costly in this case, moving and depositing $2N$ particles instead of $N(1+1/Z)$, the great advantage is that the statistical noise associated with the ion number is reduced.}

In the homogeneous case, the plasma in $xy-$plane is an ellipse with uniform density $n_{\max}$ and minor axis $2R_{\rm cx}$ along the $x-$direction, and major axis $2R_{\rm cy}$  along $y-$ direction. For inhomogeneous case, the plasma density distribution is $n =n_{\max}\exp(-x^2/w_{\rm x}^2)\exp(-y^2/w_{\rm y}^2)$, where  $w_{\rm x:y}$ is width along the $x-$ or $y-$direction. We injected from the $z_{\min}$ boundary a linearly $x-$polarized Gaussian pulse propagating in the positive $z-$direction. We applied a phase $\phi(r)=-(2\pi/\lambda) r\sin\theta$ to the Gaussian beam to create a Bessel-Gauss beam.\cite{Ardaneh_20} The Bessel beam length for this setup is $\approx$ 18~\textmu m. The peak intensity in the Bessel zone is $6\times 10^{14}\,{\rm W/cm^{2}}$ in absence of plasma (pulse energy  1.2 \textmu J). 

\begin{table}[!htb]
\centering
\begin{threeparttable}
\caption{Simulation parameters.\label{dsh}}
\begin{tabular}{ |p{4cm}||p{4cm}|  }
 \hline
Parameter & Value \\
 \hline
Computational box    & $15\times15\times 30$~\textmu $\rm m^3$   \\
Spatial resolution     &   $\Delta_{\rm x:y}k^{0}_{\rm r}=0.04$\tnote{a},  $\Delta_{\rm z}k^{0}_{\rm z}=0.1$\tnote{b} \\
 FDTD scheme  & Second-order  \\
 Boundary condition (fields) &  Perfectly matched layers\\
 Boundary condition (particles) &  Outflow\\
 Pulse energy &1.2 \textmu J \\
 Central wavelength ($\lambda_0$)  & 0.8~\textmu m\\
Cone angle ($\theta$ in air)  & $25 ^{\circ}$\\
 Pulse temporal profile & $\exp[-(t-t_{\rm c})^2/T^2]$\\
 Central time ($t_{\rm c}$) & $130\,{\rm fs}$ \\
 Pulse ${\rm FWHM}=\sqrt{2\ln2}T$    & $100\,{\rm fs}$\\
 Pulse spatial profile &  $\exp(-r^2/w_0^2)$ \\
 Pulse spatial waist ($w_0$) & 10~\textmu m \\
 Maximum density ($n_{{\max}}$) & $5-10\,n_{\rm c}$\\
 Density profile  (axial) & $\tanh(z/z_0)$\tnote{c} \\
 Critical radius ($R_{\rm cy}$) & $70-500$ nm \\
 Critical radius ($R_{\rm cx}$)  & $25-200$ nm\\
Mass ratio ($m_{\rm i}/m_{\rm e}$) & $102\times1836$\tnote{d}\\
Particle distribution [$f(\mathbf{v})$]  & Maxwellian\\
Electron temperature ($T_{\rm e}$) & $1\times10^{-3}-10$~eV\\
Ion temperature ($T_{\rm i}$) & $1\times10^{-3}-10$~eV\\
 Particles per cell per species &  32\\
Particles weight profile &  triangle\\
Run time ($t_{\rm run}$) & $320\,{\rm fs}$\\
 \hline
\end{tabular}
\begin{tablenotes}
\item [a] $k^{0}_{\rm r}=k_0\sin\theta$. \item [b] $k^{0}_{\rm z}=k_0\cos\theta$. \item [c] $z_0=1$~\textmu m. \item [d] 102 is the sapphire molar mass. 
\end{tablenotes}
\end{threeparttable}
\end{table}

{ All simulations are run twice, first without the collisions, and then with binary collisions.} The order of magnitude of the electron temperature $T_{\rm e}$, in eV, can be estimated from: \cite{kruer_1988}

\begin{equation}\label{nuc}
\nu_{\rm ei}=3\times10^{-6}\ln\Lambda\frac{nZ}{T_{\rm e}^{3/2}}=f(Z, n, T_{\rm e})
\end{equation}
where $\Lambda$ is the Coulomb logarithm, $Z$ ionization degree, and $n$ in cm$^{-3}$. { The Coulomb logarithm depends on temperature as $\Lambda\approx9  N_{\rm D}/Z$,  where the number of electrons in Debye sphere is $N_{\rm D}=1.7\times10^9 \sqrt{T_{\rm e}^3 / n}$.\cite{kruer_1988} For the given $(n,\nu_{\rm ei})$ in Fig. \ref{nu_tem}, and $Z=1$,  we have solved the nonlinear equation $\nu_{\rm ei}-f(Z, n, T_{\rm e})=0$ for $T_{\rm e}$ with the iteration tolerance equal to $1\times10^{-10}$.} We have obtained an electron temperature in the range of $\approx 5-10\, {\rm eV}$ corresponding to the range of the effective damping frequency and electron density in Fig. (\ref{nu_tem}). Our results were not affected significantly for the initial electron and ion temperatures in a range 10$^{-3}$ to 10~eV, and $n_{\max}$ in the range of $5-10\,n_{\rm c}$.
We have also verified the energy conservation by performing simulations with different numbers of particles per cell in the range of 32-128.  We have also verified the results using different FDTD schemes and particle shapes. The fourth-order schemes did not give any notable difference in the results. We achieved similar results applying a third-order b-spline particle shape with 5 points (fifth-order particle weighting). For our simulations, the numerical heating grows as ${\rm d}T_{\rm eV}/{\rm d}t \sim 1\,{\rm eV/ps}$ and is therefore negligible over the 320~fs of the simulation time in comparison with the particle heating by the laser pulse. \cite{Arber_2015} { We note that the simulation time is much smaller than the time scale of energy relaxation between the electrons and ions given by $\tau^{\rm{E}}_{\rm{ei}} \approx {1}/{Z}\left({m_{\rm{i}}}/{m_{\rm{e}}}\right) \tau^{\rm p}_{\rm{ei}}$,\cite{eliezer_2002} where the momentum loss time $\tau^{\rm p}_{\rm{ei}}\approx1$~fs [Eq. (\ref{nuc})].}

\section {Results} \label{Results} 

\subsection{Field structure}

\begin{figure*}[!htb]
	\centering
		\includegraphics[width=0.99\textwidth]{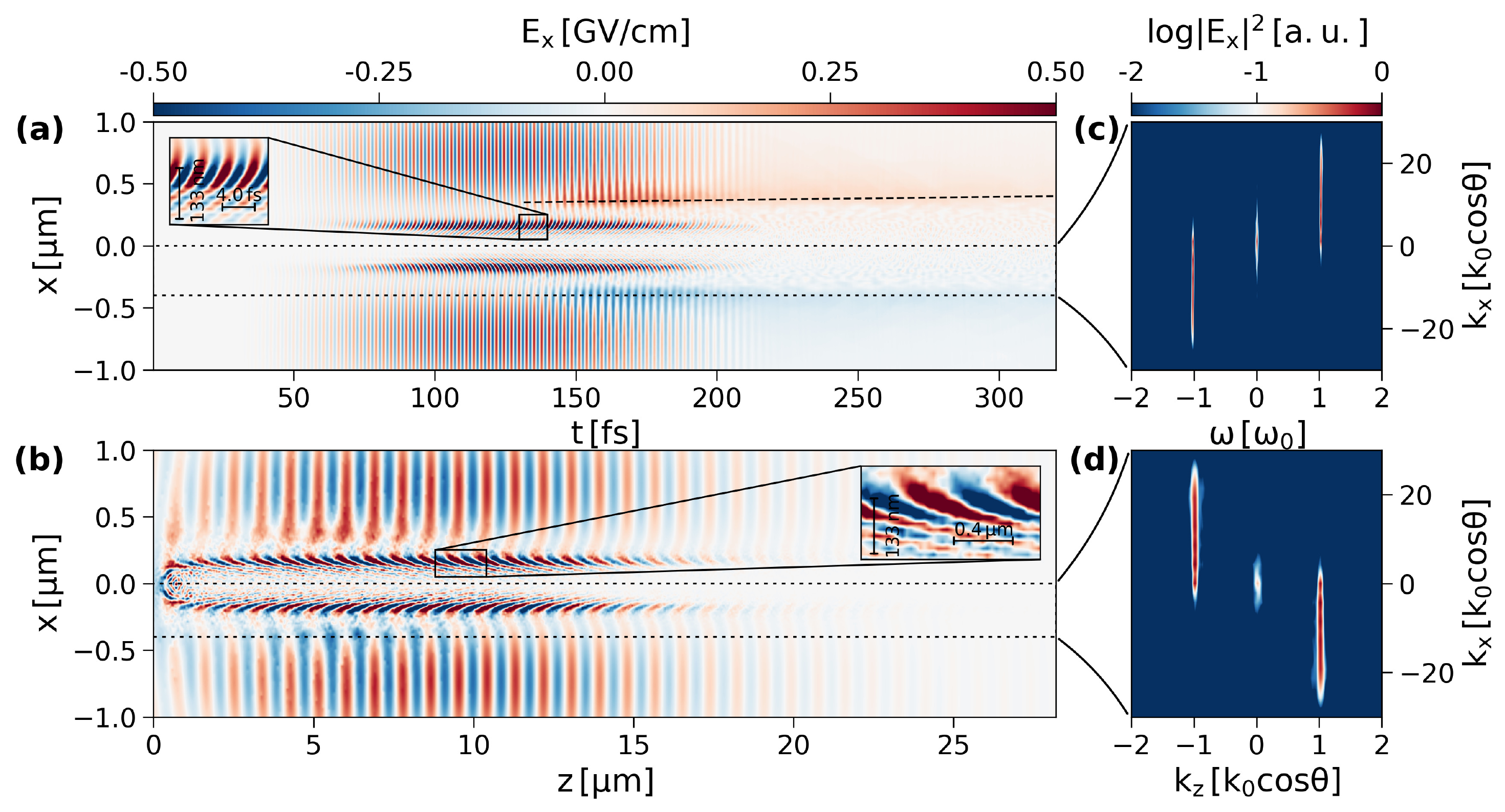}
	\caption{Inhomogeneous simulation. The $x-$component of the electric field shown in the $xt-$space, panels (a), and in the $xz-$space { at $t=150$ fs}, panels (b). The Fourier transforms performed for the dotted boxes are shown in panels (c), (d) with the log-scale color-bar.  The dashed line in the panel (a) shows plasma expansion at a speed of $\approx 1\times10^7\,{\rm cm/s}$. The insets in panels (a) and (b) show zoom-in of the electric field.}
	\label{NearField_Ex_RA}
\end{figure*}

\begin{figure*}[!htb]
	\centering
		\includegraphics[width=1\textwidth]{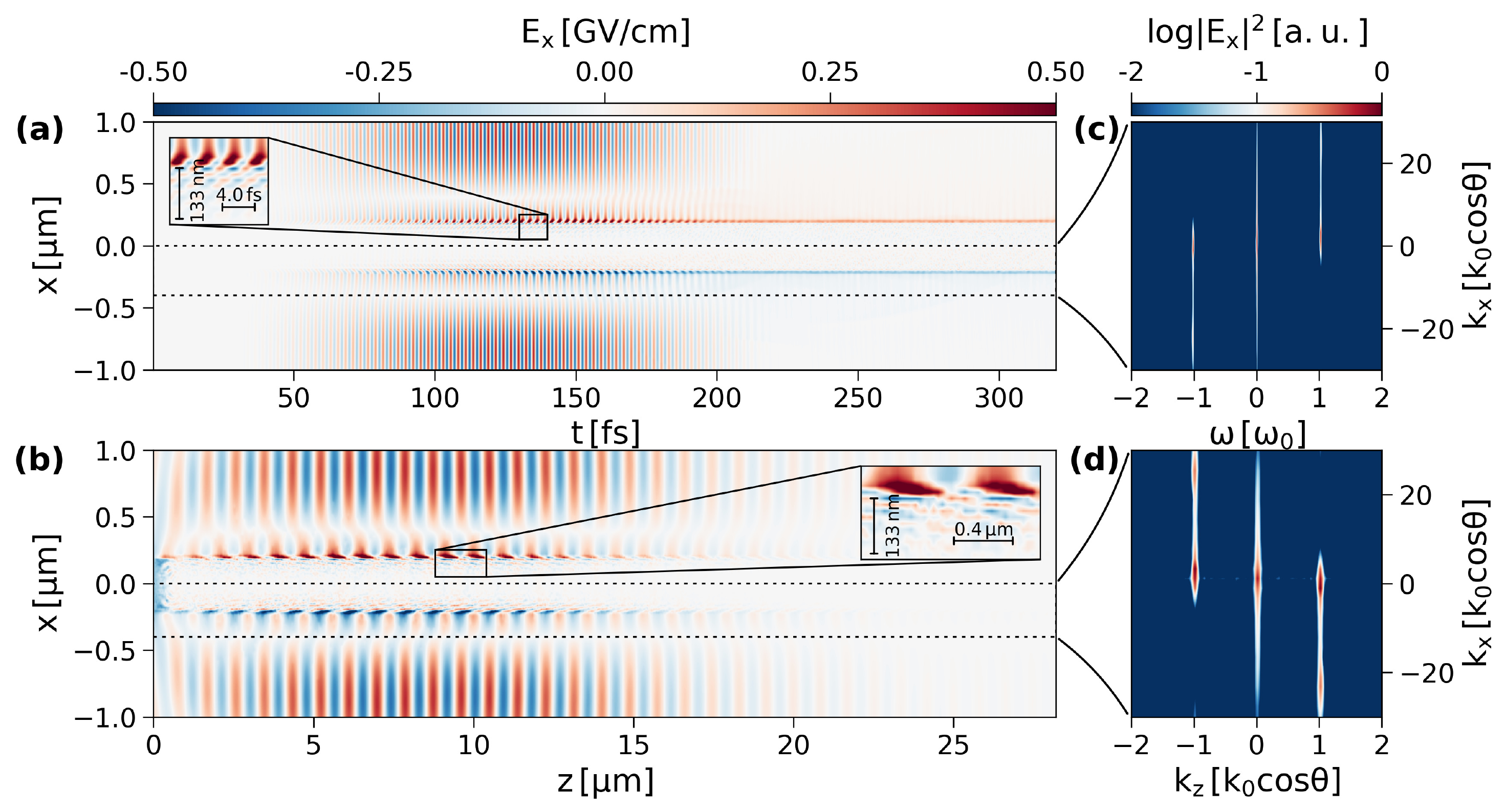}
	\caption{Homogeneous simulation. The $x-$component of the electric field shown in the $xt-$space, panels (a), and in the $xz-$space { at $t=150$ fs}, panels (b). The Fourier transforms performed for the dotted boxes are shown in panels (c), (d) with a log-scale color-bar. The insets in panels (a) and (b) show zoom-in of the electric field.}
	\label{NearField_Ex_SP}
\end{figure*}

We show in Fig. \ref{NearField_Ex_RA} the field structure for the inhomogeneous plasma simulation. One can see the $x-$component of the electric field, in the direction of the pulse polarization,  in $xt-$space in panel (a), and in $xz-$space in panel (b). One can see the excitation of electron plasma waves and resonance at the critical surfaces, at $x=\pm 190$~nm.  The time evolution of the plasma waves intensity follows the temporal profile of input laser intensity. In panels (c), and (d), we show, in log scale, the Fourier transforms for the dotted boxes of panels (a) and (b). They indicate the formation of electron plasma waves at  $(\omega,k_{\rm z})=(\omega_{0},k_{\rm 0}\cos\theta)$.  

In the sub-critical sides of the density profile, the dispersion relation for plasma waves is given by the Bohm-Gross relation $3k_{\rm x}^2v^2_{\rm th}/\omega^2 = 1-n/n_{\rm c}$, where $v_{\rm th}$ denotes the electron thermal velocity.\cite{kruer_1988,Goldston_1995} Because of the strong temperature and density gradients in the sub-wavelength plasma rod (see Sec. \ref{Electron heating} ), these waves have a range of wavevectors $k_{\rm x}$. In the over-critical side, one can see the propagation of  electron sound waves. These waves can be clearly observed in Fig.~\ref{NearField_Ex_RA}(b). The dispersion relation for these waves in collisionless plasma is $\omega\approx 1.35 k_{\rm x} v_{\rm th}$ (see Appendix \ref{Dispersion relation for longitudinal waves}).\cite{Holloway_1991,Goldston_1995} In contrast to a sound wave in a gas, these waves are mediated both by the electric field and the pressure gradient.  The greatest difference comes from collisionless kinetic effects associated with the group of particles that move at velocities close to the wave phase velocity.  This results in strong damping of the electron sound waves.  The strong gradients generate a range of $k_{\rm x}$ wavevectors. The range of $k_{\rm x}> k_0$ in these two regions is observed in Figs.~\ref{NearField_Ex_RA}(c) and (d).  

In terms of damping,  the plasma waves are mainly Landau damped in the sub-critical sides of the density profile.  The decay of the electromagnetic wave from the turning point up to the critical surface is given by $\exp(-2q_{\rm p}^{3/2}/3)$ where $q_{\rm p}=(\omega_0 L/c)^{2/3}(\omega/\omega_0)^{4/3}\sin^2 i$ is the modified Denisov absorption parameter for pulses.\cite{Palastro_2018} For $L\approx 60\,{\rm nm}$, $i=\pi/2-\theta=65^{\circ}$, and ultrashort pulses in our simulation $q_{\rm p}\approx0.5$,  the decay scale is $2L\sin^3 i/3\cos i \approx 70\,{\rm nm}$. This is consistent with the width of the resonance field at FWHM $\approx 50\,{\rm nm}$ in the simulation.

We also see in Fig.~\ref{NearField_Ex_RA}(a) and (b) that ambipolar electrostatic fields develop at the surface of the plasma starting from $t\approx t_{\rm c}$. The ambipolar fields, on either side of the plasma, are directed outward of the plasma surface parallel to $x-$direction. These fields are caused by the pressure difference between the electrons and ions. The radiation pressure of resonantly driven fields causes fast expansion of electrons relative to the ions and consequently a charge separation. The expansion is at a speed of $\approx 1\times10^7\,{\rm cm/s}$, shown by the black dashed line in Fig. \ref{NearField_Ex_RA}(a), which is roughly the sound speed $c_{\rm s}=( 3T_{\rm e}/m_{\rm i})^{1/2}$ for the temperature of hot electrons $\approx100\,{\rm eV}$ in the simulation (see Sec. \ref{Electron heating}). The amplitude of the ambipolar field reaches $\approx 1{\rm GV/cm}$.  When the laser is off, $t>230\, {\rm fs}$, the ambipolar electric fields decay with the electron-ion energy exchange rate which is in the order of the magnitude of several ps. The electrostatic ambipolar field is visible in the Fourier transforms as the bright spots at the center, $(\omega,k_{\rm x},k_{\rm z})=(0,0,0)$.

We show the results for the homogeneous simulation in Fig. \ref{NearField_Ex_SP}.  For this case, the waves are confined to the surface of the plasma as one can see in panels (a), and (b).  As discussed in Sec. \ref{Surface plasmon}, these surface waves are mainly electromagnetic. Therefore, in the Fourier transforms in panels (c) and (d), the surface waves are identified at $(\omega,k_{\rm x},k_{\rm z})\approx(\omega_{0},0,k_0\cos\theta)$. The surface waves heat the electrons in the vicinity of the surface. The heated electrons then expand into the surrounding medium and leave positive charge at the surface of the plasma. As a result, an electrostatic component is generated which is visible in the center of Fourier transforms, $(\omega,k_{\rm x},k_{\rm z})=(0,0,0)$.  The electron sound waves are also present inside the plasma. These waves are clear in the Fourier transforms at $k_{\rm x}\gg k_0$. 

The surface waves have the maxima at the surface and  decay on both sides of the surface while damping in the plasma side is much stronger. For a plasma with density $n = 5n_{\rm c}$ and at temperature 1~eV, which is the main component of the electron temperature in simulation (see Sec. \ref{Electron heating}), we calculated a damping time of 0.35 fs ($\nu_{\rm eff}=1.2 \omega_0$) using binary collisions, in agreement with Eq.~(\ref{nuc}). The plasma permittivity  is therefore $\epsilon_{1}=-1.1+j2.5$ [Eq. (\ref{eps_plasma})]. The decay scale of the surface waves is $1/\vert k_{x2}\vert\approx 36\,{\rm nm}$ in the medium ($\epsilon_2=1$ in the simulation because the plasma is in contact with air) and $1/\vert k_{x1}\vert\approx 33\,{\rm nm}$ in the plasma [Eq. (\ref{SPE}b)], almost half of the damping length in the inhomogeneous plasma. The decay scale is compatible with the width of the surface wave at FWHM $\approx 50\,{\rm nm}$ in the simulation.  The excitation of the surface wave in the interface of a free damping plasma and air requires grating coupling to bridge the momentum difference between the incident laser and surface plasmon. However, the presence of damping, $\nu_{\rm eff}$, and roughness of surface reduce the momentum mismatch between surface plasmon and incident light and satisfy the condition for surface wave excitation (see Appendix \ref{Dispersion relation for surface plasmons}).

In summary, we have seen resonantly driven electrostatic electron waves in the inhomogeneous plasma and electromagnetic surface waves in the homogeneous plasma. For both cases, there are ambipolar fields at the plasma surface and electron sound waves propagating into the high-density region of the plasma. 

\subsection{Electron heating}\label{Electron heating} 

\begin{figure*}[!htb]
	\centering
		\includegraphics[width=\textwidth]{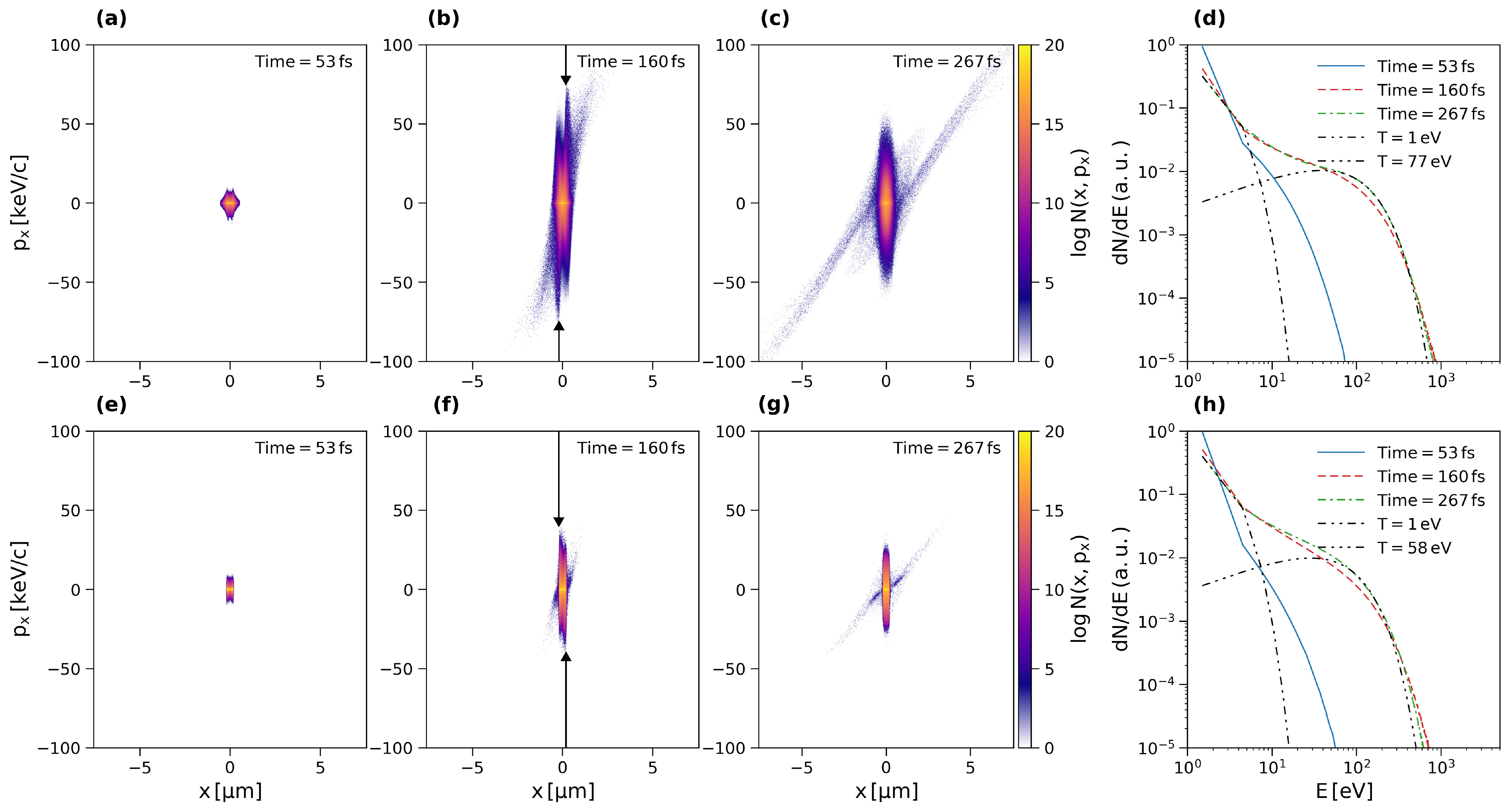}
	\caption{The electron phase-space plots at three different times for the inhomogeneous plasma, panels (a)-(c), and homogeneous plasma, panels (e)-(g). The electron energy distribution functions {at three different times} are shown in panels (d) for inhomogeneous plasma, and panel (h) for homogeneous plasma. { The black dashed-double-dotted, and black dashed-triple-dotted lines in the left column show the Maxwellian energy distribution function fits. The distribution functions are normalized by the total number of particles in the systems.} The arrows in panel (b) indicate the critical surfaces while in panel (e) indicate the plasma surface.}
	\label{disb}
\end{figure*}

We studied the interaction of electrons with electrostatic and electromagnetic fields by tracing the most energetic electrons. We found that if an electron enters the resonantly driven fields with the proper phase, it will experience a sharp energy increase. The transit acceleration for electrons happens as electrons surf the oblique structure of the resonantly driven electric field, $D\approx 70$~nm in the simulation. The duration of transit acceleration is typically half of a field oscillation \cite{Morales_1974,DeNeef_1977}, 1.3 fs in the simulation.  The maximum attainable energy for an electron during transit acceleration is $E_{\max}=\sqrt{\pi} D E_{\rm c}\approx 6\,{\rm keV}$, which corresponds to the integration of a Gaussian function $E_{\rm c}\exp\left(-x^2/D^2\right)$, where $E_{\rm c}\approx0.5\,{\rm GV/cm}$ is the amplitude of the resonantly driven electric field. This corresponds with a maximum momentum of electrons $p_{\max}\approx 75\,{\rm keV/c}$.  A small population of electrons, $\approx$~1\%, interacts with the plasma waves at the critical surface, shown with vertical arrows in Fig. \ref{disb}(b). These electrons are heated up to $p_{\rm x}\approx 75\,{\rm keV/c}$, in agreement with the transit acceleration.

The process of electron heating at the surface of homogeneous plasma acts similarly. However, the surface waves have a smaller thickness $D$ relative to the plasma wave leading to a smaller level of attainable energy for the electrons, $\approx 1.5\,{\rm keV}$ and $p_{\rm x}\approx 40\,{\rm keV/c}$. There are more particles at the surface of the plasma in interaction with surface waves. The electron temperature for the same amount of absorption is, therefore, lower in this case, Figs. \ref{disb}(f) and \ref{disb}(g). 

The electrons are heated in the interaction with plasma waves, and ambipolar electric fields in the inhomogeneous plasma case, and surface waves in the homogeneous case. Due to the heating, a single Maxwellian distribution does not fit the electron energy distribution as shown in Figs. \ref{disb}(d), and \ref{disb}(h). The fit to the electron energy distribution at the late stage consists of two Maxwellian distributions in the form of $f(E)=A\sqrt{E/T_1^3}\exp(-E/T_1)+B\sqrt{E/T_2^3}\exp(-E/T_2)$. In the fit, $T_1 =1\, {\rm eV}$ for both inhomogeneous and homogeneous plasmas, the same order of magnitude of the value presented in Sec. \ref{Theoretical constraints}. The temperature of the hot electron population is $T_2 =77\, {\rm eV}$ in the inhomogeneous case while it is $T_2 =58\, {\rm eV}$ in homogeneous case. The lower temperature for the homogeneous plasma is due to a higher number of electrons in interaction with the surface waves, and smaller width of the surface wave. 

The electrons expand toward the free space due to the radiation pressure of the plasma waves. The expanding electrons encounter the ambipolar electric fields of the double layers. In later times, the ambipolar electric field is given by $E=-{1}/{en_{\rm n}}{{\rm d}P_{\rm e}}/{{\rm d}x}$.\cite{Goldston_1995, hora_2008} Therefore, the work required for moving an electron from a high-density $n_1$ to a low-density $n_2$, considering an adiabatic equation of state, is $-e\Delta\phi=\gamma T_{\rm e}\ln(n_1/n_2)\approx4\,{\rm keV}$. The energetic electrons with energies $E\gtrsim-e\Delta\phi$ will escape the plasma volume. These electrons are visible as two counter-streaming outflows indicated by arrows in Fig. \ref{disb}b.   In the homogeneous plasma case, the electrons are less heated and the outflows are weaker than the inhomogeneous plasma. 

\subsection{Comparison with experiments}
We have compared the measurable quantities from the experiments with the results from simulations. It includes a comparison of the near-field fluence and far-field radiation pattern.

\subsubsection{Near-field}

\begin{figure*}[!htb]
	\centering
		\includegraphics[width=\textwidth]{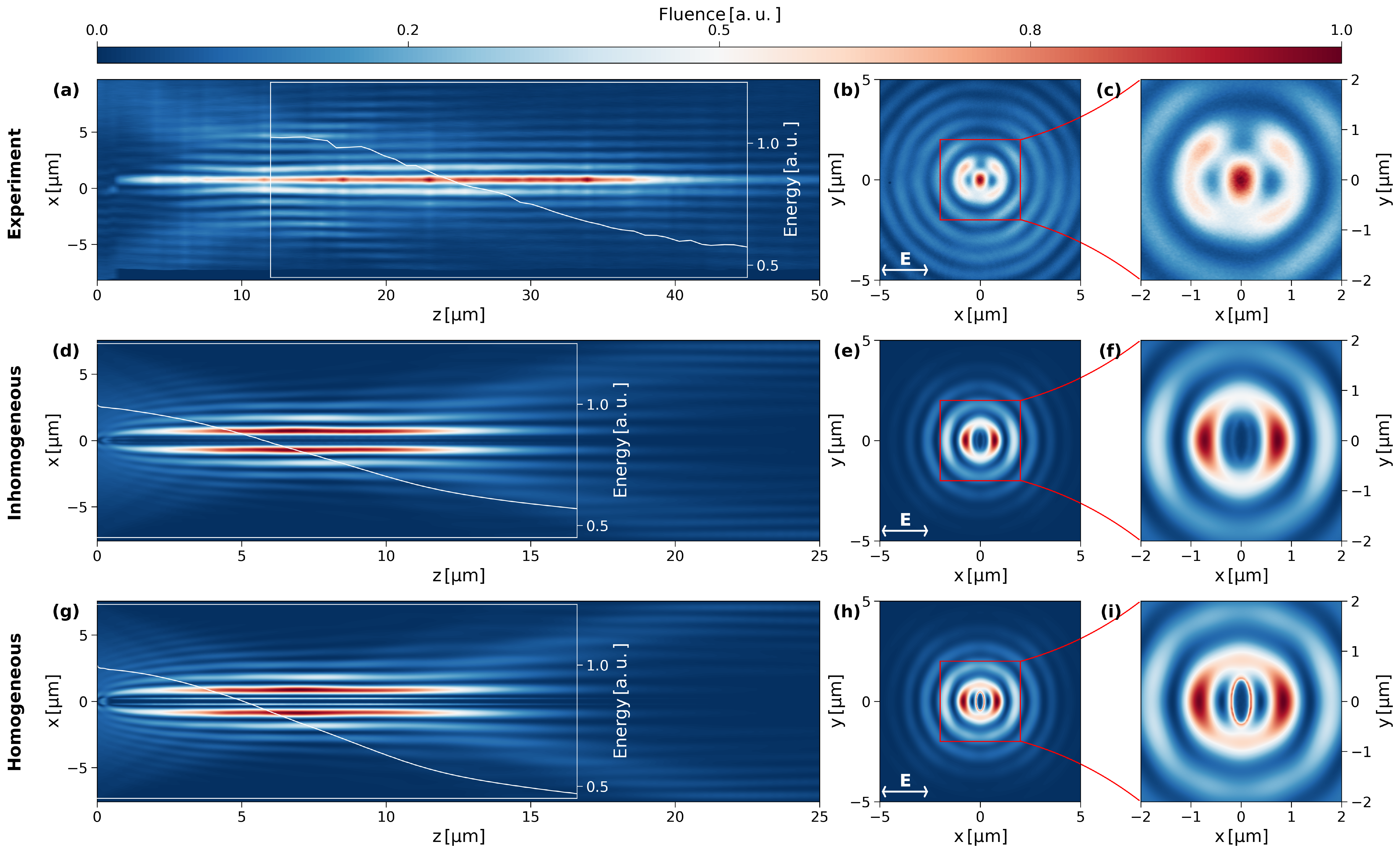}
		\caption{Near-field fluence maps for experiments in sapphire (top row) and simulations (inhomogeneous plasma  profile in middle row and homogeneous plasma  profile in bottom row). Panels (a), (d), and (g) are slices of the fluence in the $zx-$plane ($y=0$).  Panels (b), (e), and (h) are slices in $xy-$plane at $z=20$~\textmu m in (b) and $z=7$~\textmu m in (e) and (h). Panels (c), (f), and (i) are zoom-in of (b), (e), and (h), respectively. The over-plotted lines in (a), (d), and (g) show the pulse energy as a function of propagation distance $z$. {We have removed the electrostatic field components in the fluence distribution calculations.}}
	\label{NearField}
\end{figure*}

Figure \ref{NearField} shows the slices of fluence $S_{\rm z}$, {\it i.e.} ({time integration of intensity $cB^2/8\pi$, instead of Poynting vector in $z-$direction, to eliminate the contribution of electrostatic waves}) for the experiments (top row), and simulations (middle and bottom rows). As the electrostatic structure cannot be observed experimentally by beam imaging, we exclude from the fluence calculation the electrostatic field components. In all cases, the input laser pulse is linearly polarized along the $x-$direction. The middle row corresponds to irradiation of an elliptical plasma rod with an inhomogeneous density while the bottom row corresponds to an elliptical plasma rod with a homogeneous density.
For these simulations, the critical radius has been set to $R_{\rm cx}=190$~nm along the $x-$direction (polarization direction) and $R_{\rm cy}=450$~nm along the $y-$direction. The insets in panels (a), (d), and (g) represent the pulse energy, determined by surface integral of the fluence $S_{\rm z}$ over the $xy-$surface, as a function of the propagation distance. As one can see, there is a linear decrease of the pulse energy over a length of $\approx 18$~\textmu m with a slope of $\approx-0.03$~\textmu J/\textmu m in both experiments and simulations. In the experiments, this length corresponds to the axial length of the generated plasma. The overall absorption factor is approximately 0.5 in both experiments and simulations. There is good agreement between the experiments and simulations regarding the absorption factor and slope of the energy curve for both homogeneous and inhomogeneous profiles. The figures show the results for collisional simulations. We have observed that the structure of the fields in the collisionless simulations is almost identical to the collisional one. The difference in terms of absorption between the collisional and collisionless simulations was around 10 percent. Therefore, the main mechanism of energy deposition is collisionless in both homogeneous and inhomogeneous cases and relies on electron-wave interactions.

The near field fluence distribution of the Bessel beam in vacuum is cylindrically symmetric.\cite{Ardaneh_20} This is not the case in our experiments and simulations, as apparent in Figs. \ref{NearField}(b) and \ref{NearField}(c). In a preceding work,\cite{ardaneh_2021} we investigated the origin of this asymmetry by comparing numerical simulations for different plasma parameters. We found that the asymmetry is associated with the plasma ellipticity and orientation with respect to the laser polarization.  

The presence of a plasma with a density larger than $ n_{\rm c}\cos^2 i$, {\it i.e.}, the turning point density, defocuses the input laser beam due to the reflection at the turning point. As a result, we observe a very small outward shift of the circular lobes and a depletion of the intensity at the center. The lobe shift is $\approx$ 300 nm in our simulations, Figs. \ref{NearField}(f) and \ref{NearField}(i). Importantly, the central hole observed in our simulations is filtered in the experiments by the imaging operation. { We noted that the imaging operates at the exit side of the sample (see Fig. \ref{exp_setup}). Therefore, the fine details of the fields inside the central lobe are not transferred to the imaging plane.}

Despite similar absorption, the slope of the energy curve, and outer lobes, the inhomogeneous and homogeneous plasmas show two different field structures within the plasma region. { In the case of the homogeneous plasma, Fig. \ref{NearField}(i), there is an electromagnetic structure (elliptical ring) at the surface of the plasma because the surface waves are electromagnetic [Figs. \ref{NearField_Ex_SP}(c) and \ref{NearField_Ex_SP}(d)].  However, the structures inside the inhomogeneous plasma in Fig. \ref{NearField}(f) are electrostatic and hence not captured in the fluence pattern (time integration of intensity $cB^2/8\pi$).}  Both the surface waves and electron plasma waves are bound modes. It means that they are absent in the measurements. 

\subsubsection{Far-field}
In our simulations, we calculated the far-field intensity as follows. We recorded field data at a fixed propagation distance of $z=20$~\textmu m for $\vert x \vert \leqslant 5$~\textmu m and $\vert y \vert \leqslant 5$~\textmu m. We have computed the intensity spectrum $I(\omega,k_{\rm x},k_{\rm y})$ by performing a discrete Fourier transform on each component of the magnetic field, $B_{\rm x:y:z}(t,x,y)$. We then filtered the intensity spectrum at the central frequency of the pulse $\omega_0$ for comparison with the experimental far-field intensity map.

\begin{figure}[!htb]
	\centering
		\includegraphics[width=\columnwidth]{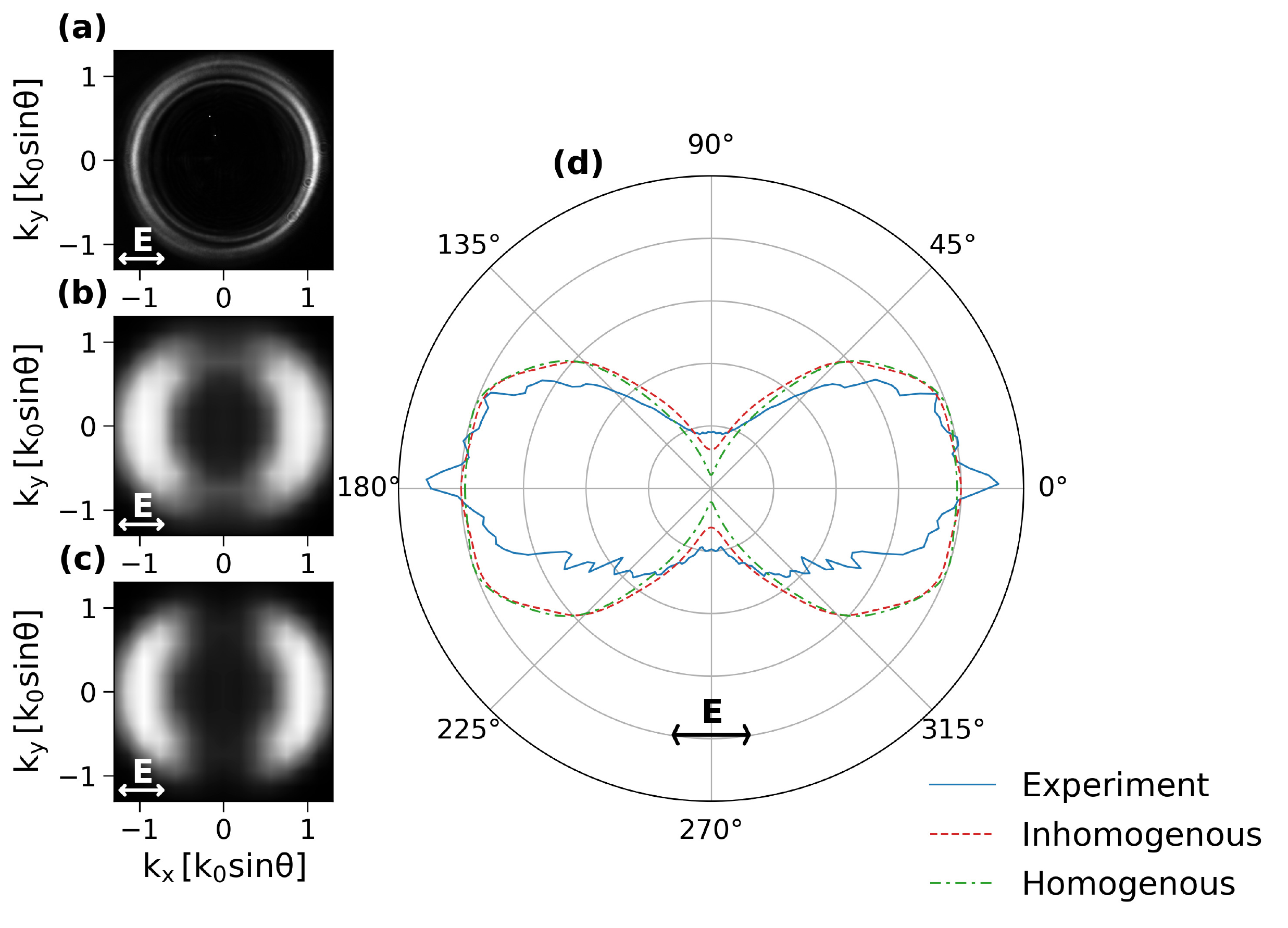}
	\caption{The far-field radiation at $\omega_0$ in the experiment, panel (a), and simulations: panel (b) for inhomogeneous plasma, and panel (c) for homogenous plasma. The angular distribution for each distribution is shown in panels (d). }
	\label{FarField}
\end{figure}

The far-field experimental distribution is shown in Fig.~\ref{FarField}(a). It is composed of two bright lobes at $k_0\sin\theta$ parallel to the input laser polarization. A significant part of the pulse energy is absorbed through the resonance of the electron plasma waves or surface waves parallel to the laser polarization. Therefore, for a circular plasma rod, the far-field radiation pattern consists of two parallel lobes perpendicular to the laser polarization. However, this is not the case for our experiments.  In a parametric study of the far-fields for different plasmas, we found that the plasma shape orientation with respect to the laser polarization has a significant impact on the far-field pattern.\cite{ardaneh_2021} Similar to the near-field fluence, we could only reproduce the experimental far-fields using elliptical plasma rods elongated as the major axis is perpendicular to the input laser polarization as we used in this work (Table \ref{dsh}).

As shown in Figs.~\ref{FarField}(b) and ~\ref{FarField}(c), the far-field radiation patterns for both inhomogeneous and homogeneous plasma well match the experimental ones. The distributions from the simulations are broader relative to the experiments that is due to performing the Fourier transform on a limited window. The angular distribution in Fig.~\ref{FarField}(d) represents a quantitative agreement between the simulations and experiments. The radiation pattern intensity at $90^{\circ}$, poles of the ellipse, is at the minimum and is fainter for the homogeneous plasma.

\section{Discussion}\label{Discussion}
Ultrafast femtosecond Bessel beams can create sub-wavelength over-critical plasmas inside dielectrics. It is confirmed by comparing the measurements with the results from the ab-initio PIC simulations for diagnostics including near-field distribution, absorption, and far-field radiation patterns for the sapphire. Here we have demonstrated that the formation of surface waves or resonance of electron plasma waves in elliptical plasmas can mediate the deposition of laser energy. In the former scenario, electromagnetic waves develop at the surface of an over-critical plasma due to the permittivity difference between the plasma and surrounding material. In the latter scenario, electron plasma waves form in an inhomogeneous plasma by the electric field component of the laser parallel to the density gradient. Both scenarios showed a good agreement with our measurements.  We remark that in the homogeneous case, they remain more in the vicinity of the plasma surface than in the inhomogeneous case. We expect similar absorption processes in other dielectrics such as fused silica and glass. The suggested processes are collisionless and rely on wave-particle interactions. These processes might also explain the high absorption required for void formation in fused silica using Bessel beams, even for a shorter duration of 50 fs reported by Bhuyan {\it et al} \cite{Bhuyan2017} and Beuton {\it et al}. \cite{Beuton_2021} 

Surface waves and electron plasma waves are two limits for high absorption in the sub-wavelength plasma.  A surface wave is an electromagnetic non-propagating feature and forms due to the difference in the permittivity of two neighboring dielectrics. The plasma wave, in contrast, is electrostatic and form as the result of the evanescence field beyond the turning point and the presence of a density gradient. The surface wave is the {limiting} case of the plasma wave when the slope of the density gradient approaches infinity. As there are more electrons in the vicinity of the surface waves, the final temperature of homogeneous plasma would be smaller than the inhomogeneous one, for the same absorbed energy. Moreover, due to the smaller width of the surface waves, the in-phase electrons with a wave surf shorter time on wave resulting in lower attainable energy for electrons. 

The electrons that are in phase with surface waves or plasma waves are heated efficiently. The heated electrons expand into the surrounding medium and leave the positive ions at the plasma surface. It leads to the formation of ambipolar electric fields at the plasma surface.  Due to the sharper pressure gradient, the potential at the surface of homogeneous plasma is stiffer and a smaller population of electrons can escape from it.  In addition, the propagation of the hot electrons into over-critical plasma develops the electron sound waves. At $kv_{\rm th}/\omega_{\rm pe}\gg 1$ (short wavelength), or high temperature, the Bohm-Gross dispersion relation commences looking like an electron sound wave. The phase and group velocities both converge to $\sqrt{3}v_{\rm th}$.  The damping of these waves is as large as their frequency because there is a large class of particles that move at velocities close to the wave phase velocity.  

It is possible that the difference in plasma heating between the two scenarios will explain part of the morphological differences observed during material modifications with Bessel beams of various laser pulse parameters. 
We anticipate that measurement of the second harmonic can provide further insights into the absorption mechanisms at play during the interaction of femtosecond pulses with solid dielectrics. In a forthcoming work, Ardaneh {\it et al}\cite{ardaneh_2022}, we will explicitly compare the second harmonic generation from the experiments with the PIC simulations. Further work will be dedicated to simulations of the ionization dynamics within PIC simulations to provide the plasma evolution  during the interaction of femtosecond pulses with solid dielectrics. 

\section{Conclusions}\label{Conclusions}
{ We have estimated that focusing 100 fs (FWHM) Bessel beams of intensities in the order of $\sim 10^{14}\,{\rm W/cm^{2}}$ inside sapphire will generate an over-critical plasma (density between $10n_{\rm c}\lesssim n\lesssim 70n_{\rm c}$) of short scale length on the order of $L/\lambda_0\approx 0.025-0.075$. The resonance of electron plasma waves at the critical surface or the formation of surface plasmons plays an important role in the interaction between femtosecond Bessel beams with plasma inside dielectrics. These two scenarios account for more than half of the pulse energy deposition inside the dielectrics. One distinction between the results presented in this work and those in the literature is the collisionless process of absorption that relies on the wave-particle interaction. Despite the different nature of these two processes, their near-field and far-files properties match well with the experiments.}

\begin{acknowledgments}
Technical assistance by C. Billet and E. Dordor as well as fruitful discussions with J.M. Dudley and D. Brunner are gratefully acknowledged. We thank the EPOCH support team for help  \url {https://cfsa-pmw.warwick.ac.uk}, and French RENATECH network. The authors acknowledge the financial supports of: European Research Council (ERC) 682032-PULSAR, Region Bourgogne-Franche-Comte and Agence Nationale de la Recherche (EQUIPEX+ SMARTLIGHT platform ANR-21-ESRE-0040), Labex ACTION ANR-11-LABX-0001-01, I-SITE BFC project (contract ANR-15-IDEX-0003), and the EIPHI Graduate School ANR-17-EURE-0002. This work was granted access to the PRACE HPC resources MARCONI-KNL,  MARCONI-M100, and GALILEO at CINECA, Casalecchio di Reno, Italy, under the Project "PULSARPIC" (PRA19\_4980), PRACE HPC resource Joliot-Curie Rome at TGCC, CEA, France under the Project "PULSARPIC" (RA5614), HPC resource Joliot-Curie Rome/SKL/KNL at TGCC, CEA, France under the projects A0070511001 and A0090511001, and M\'{e}socentre de Calcul de Franche-Comt\'{e}. 
\end{acknowledgments}

\appendix
\section{Dispersion relation for longitudinal waves}\label{Dispersion relation for longitudinal waves}
The dispersion relation for longitudinal waves in an unmagnetized plasma reads:\cite{galeev_1989}

\begin{equation}\label{ESDR}
D\left(k, \omega+j \nu_{\rm c}\right)=1+\frac{\omega_{\mathrm{\rm pe}}^{2}}{k^{2}} \int \mathrm{d}^{3} \mathbf{v} \frac{\mathbf{k} \cdot \partial F / \partial {\mathbf v}}{\omega-\mathbf{k} \cdot \mathbf{v}+j \nu_{\rm c}}=0
\end{equation}

The collisional damping is introduced by $\nu_{\rm c}$. Let us assume that the wave propagate in $x-$direction ($\mathbf{k}={\mathbf{\hat{x}}}k$) and electrons have a Maxwellian distribution function at the temperature $T$.

\begin{equation}\label{EDF}
F(v)=\left(\frac{m}{2 \pi T}\right)^{3 / 2} \exp \left(-\frac{mv^2}{2 T}\right)
\end{equation}
The integration of dispersion relation gives:\cite{galeev_1989}

\begin{equation}\label{EDF1}
D\left(k, \omega+j \nu_{\rm c}\right)=1-\frac{Z^{\prime}(\zeta)}{2 k^{2} \lambda_{\rm D}^{2}} = 0
\end{equation}
where 
\begin{subequations}\label{EDR2}
\begin{align}
\zeta = &\frac{\omega+j\nu_{\rm c}}{k v_{\rm th} \sqrt{2}}\\
v_{\rm th} =& \sqrt{\frac{T}{m}}\\
Z^{\prime}(\zeta)=&-\pi^{-1 / 2} \int_{-\infty}^{\infty} {\rm d}t\frac{2 t}{t-\zeta} \mathrm{e}^{-t^{2}}=-2(1+\zeta Z)\\
Z(\zeta)=&\pi^{-1 / 2} \int_{-\infty}^{\infty} {\rm d} t\frac{\mathrm{e}^{-t^{2}}}{t-\zeta}
\end{align}
\end{subequations}

\begin{figure}[!htb]
\begin{center}
\includegraphics[width=\columnwidth]{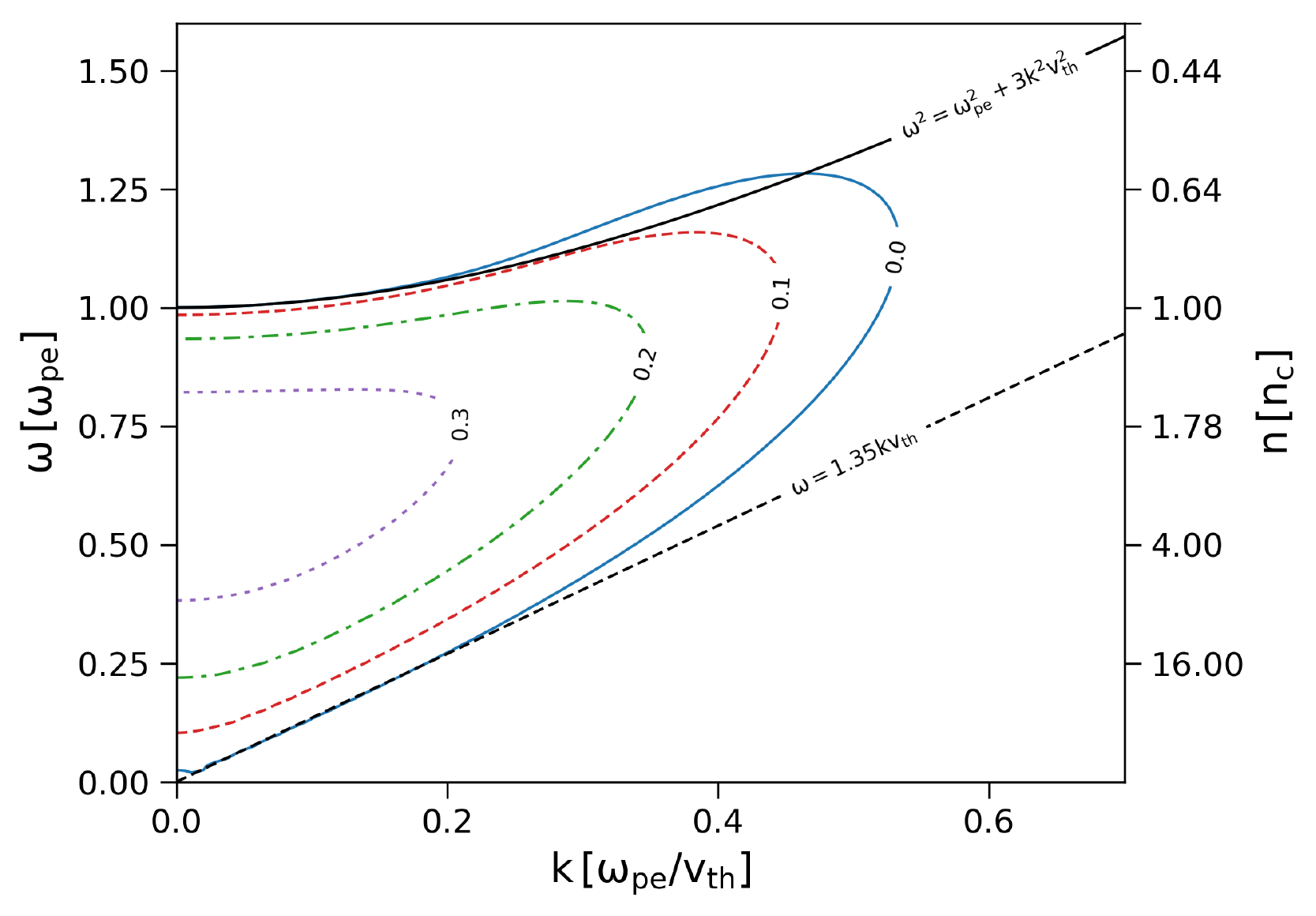}
\caption{Dispersion relation for electrostatic wave in an unmagnetized plasma for different collision frequency, normalized to $\omega_{\rm p}$. The collision frequency is indicated inline. The black solid line shows the dispersion relation for Langmuir waves while the black dashed line shows the dispersion relation for electron sound waves.}
\label{ESDRF}
\end{center}
\end{figure}

We solved Eq. (\ref{EDF1}) in  $(\omega,k)$ space for different $\nu_{\rm c}$, normalized to $\omega_{\rm pe}$. As shown in Fig. \ref{ESDRF}, the Langmuir waves and sound waves can be present in sub-critical, and over-critical plasmas, respectively. For example, the wavevector of Langmuir waves turns to zero at $n\approx1.2n_{\rm c}$ for a collision frequency of $\nu_{\rm c}=0.2\omega_{\rm pe}$, while for the sound wave  $kv_{\rm th}/\omega_{\rm pe}=k\lambda_{\rm D}\approx0.35$. For a plasma at the temperature of $T=80\ \text{eV}$, and the critical density of the $n_{\rm c}= 1.7\times10^{21}\ \text{cm}^{-3}$, the Debye length is $\lambda_{\rm D}\approx1.5\ \text{nm}$. Therefore, the wavelength of the electron sound waves in the over-critical plasma is $\lambda\approx25\ \text{nm}$. For $\lambda_0=800\ \text{nm}$ and $\theta=25^{\circ}$, $k\approx30k_0\cos\theta$.

\section{Dispersion relation for surface plasmons}\label{Dispersion relation for surface plasmons}

\begin{figure}[!htb]
\begin{center}
\includegraphics[scale=0.5]{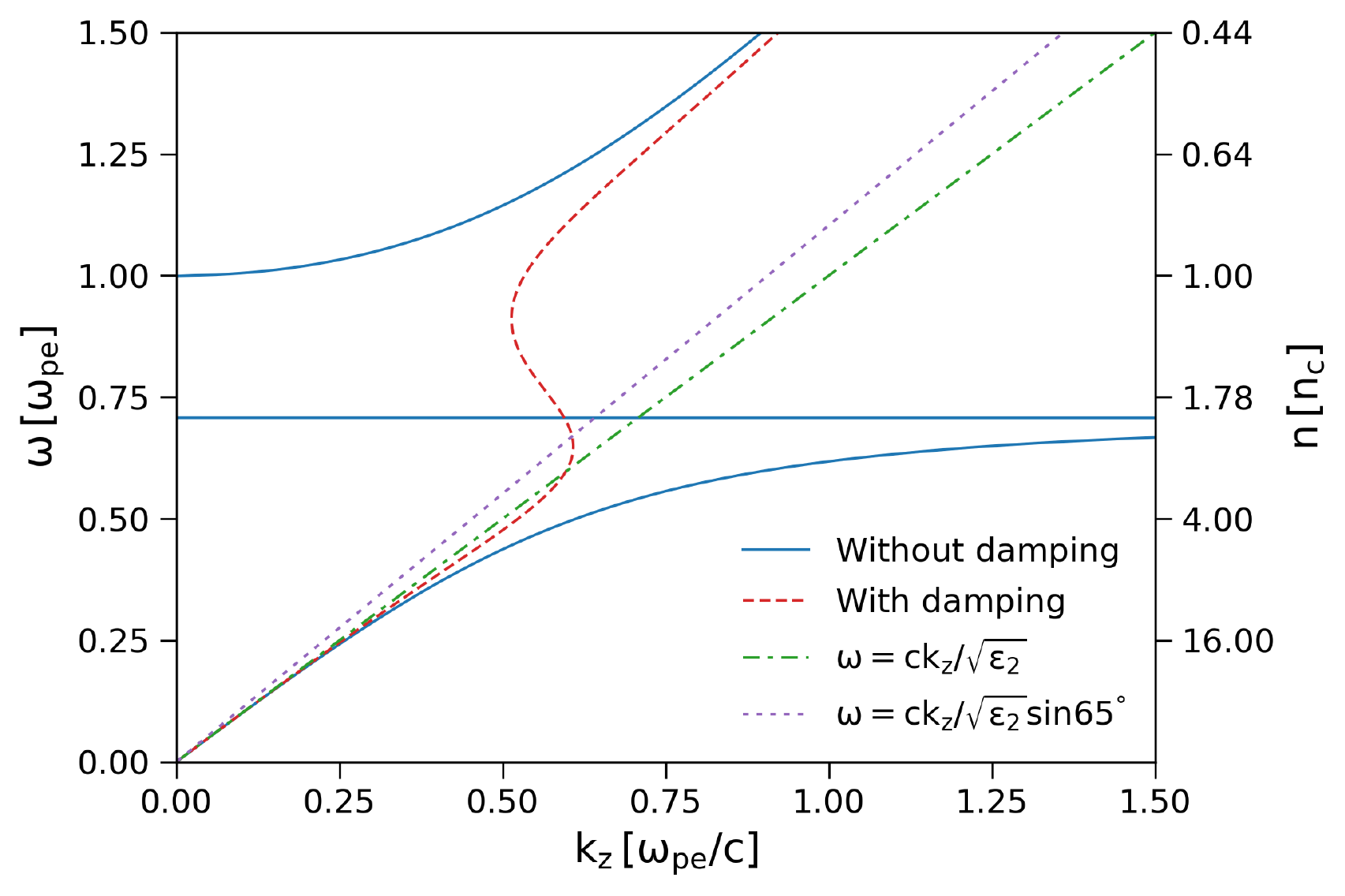}
\caption{Dispersion relation for the surface plasmon without damping (blue solid), and with damping (red dashed). The green dashed-dotted line shows the light line while the purple dotted line shows the light line along the surface for $i=65^{\circ}$.}
\label{SPF}
\end{center}
\end{figure}

In Fig. \ref{SPF}, we have shown the solutions of surface plasmon dispersion relation [Eq. (\ref{SPE}a)] for two cases, without damping $\nu_{\rm eff}=0$ and with a damping $\nu_{\rm eff}=0.5\omega_{\rm pe}$ ($\nu_{\rm eff}=1.1\omega$ for $n=5n_{\rm c}$). The surface plasmon dispersion curves in the over-critical domain stay right to the light curve, and for each frequency $\omega$, there is a small momentum mismatch in which the $k_{\rm sp}$ is slightly greater than the light's one, $k_{\rm 0}\sin i$. This figure shows that: (a) to generate surface plasmon from the light waves, the momentum difference must be provided, in some cases using grating; (b) surface plasmon has a bound or non-radiative nature; (c) the presence of damping reduces the momentum mismatch. In the range of the large $k_{\rm z}$, the frequency $\omega$ approaches to an asymptotic frequency ${\omega_{\rm pe}}/\sqrt{1+\epsilon_{2}}$. In this limit, both the group and phase velocities turn to zero and the surface plasmon shows an electrostatic localized electron oscillation. 

\section*{References}
\bibliography{manuscript}

\end{document}